\documentstyle[aps,prl] {revtex}
\twocolumn
\begin{document}
\draft
\title{Ginzburg-Landau theory of the cluster glass phase}
 \author{Th.~M.~Nieuwenhuizen$^{(1)}$ and C.~N.~A. van Duin$^{(2)}$}
 \address{
$^{(1)}$ Van der Waals-Zeeman Laboratorium, Universiteit van Amsterdam\\ 
	 Valckenierstraat 65, 1018 XE Amsterdam, The Netherlands\\
$^{(2)}$ Instituut Lorentz, Rijksuniversiteit Leiden\\
Nieuwsteeg 18, 2311 SB Leiden, The Netherlands}
 \date{email: nieuwenh@phys.uva.nl; 
cvduin@rulgm0.LeidenUniv.nl}
\maketitle
\begin{abstract}
On the basis of a recent field theory for site-disordered spin glasses
a Ginzburg-Landau free energy is proposed to describe the
low temperatures glassy phase(s) of site-disordered magnets. 
The prefactors of the cubic and dominant quartic terms change gradually
along the transition line in the concentration-temperature phase diagram.
Either of them may vanish at certain points $(c_\ast,$ $\,T_\ast)$, 
where new transition lines originate. The new phases are classified.
\end{abstract}
\narrowtext

A simple mixture of a metallic host with a magnetic atom,
such as Au$_{1-c}$Fe$_c$, is known to have a rather complicated phase diagram.
According to Mydosh ~\cite{Mydosh}  the following phases occur
at zero temperature: At very low $c$
there is  the Kondo-regime of {\it independently} 
compensated spins in a metallic host. 
At somewhat larger concentrations 
there is a spin glass phase of {\it interacting single spins} with 
$T_g \propto c$. For  $0.5\%<c<10\%$  
the spin glass experiences gradual
{\it cluster formation}, while for $10\%<c<16 \%$ one has the 
{\it cluster glass phase}. For $c>16\%$ 
one enters the percolated 
 {\it ferromagnetic phase}, which partly 
also behaves as a cluster glass.

The Kondo regime, and the low concentration spin glass phase 
are relatively well understood. The latter is described by an
Edwards-Anderson model with RKKY interactions. Its properties are
obtained from a mean field approach~\cite{ThMNsite}
and from numerical analysis, see e.g. \cite{IMISI}.
Whether or not a thermodynamic phase transition occurs in zero field
or even in non-zero field remain topics of much controversy.

Though ferromagnetism by itself is well known, clustering
properties of inhomogeneous 
ferromagnets are also far from well understood.
It is known that replica symmetry breaking
may occur before the onset of ferromagnetism,~\cite{DHSS}
possibly describing Griffiths singularities.

The situation for the clustering spin glass (with clusters containing up to
five atoms) and the cluster glass (where as many as 2000 atoms may build
a cluster; these clusters order in a glassy way) is less satisfactory.  
Little is  known about these phases.
There seems to be no experimental evidence that the given names correspond 
to thermodynamic phases that are significantly different 
from the spin glass phase. Nevertheless, the existence of
new glassy phases is the main question we wish to investigate 
theoretically in this work.

Recently one of us~\cite{ThMNsite} formulated a field theory for 
site-disordered Ising systems. With exception of the 
Kondo regime, this applies to the whole phase diagram 
of systems such as mentioned above. 
We thus consider a system with translationally invariant pair couplings 
$J(r-r')$ 
 with a fraction $c$ ($0<c<1$) of the lattice sites occupied at random.
We restrict ourselves to the second order cumulant expansion.
This description is  Gaussian in the magnetization fields, and 
equivalent to a variational (`` Hartree'') approximation.
It is quite close to the one of the SK-model since  it involves only the order 
parameters $q_{\alpha\beta}$ ($=\langle s_\alpha s_\beta\rangle$ for small
$c$) and their conjugates $p_{\alpha\beta}$.
The replicated free energy per spin reads
\begin{eqnarray}\label{freeenergy}
\beta F_n& =&  \frac{1}{2c}\int \frac{{\rm d}^3k}{(2\pi)^3} 
\sum_{\alpha}\{\ln(1-c\beta\hat{J}(k)q) \}_{\alpha\alpha}\\ 
&+&\frac{1}{2}\sum_{\alpha\beta}q_{\alpha\beta}p_{\alpha\beta}
+\sum_{l=1}^{\infty}\gamma_l\left(1-{\rm tr}_s^{(l)}\exp 
X^{(l)}\right)\nonumber
\end{eqnarray}
with $\gamma_l=(-c)^{l-1}/l(1-c)^l$ and
$X^{(l)}=\beta H\sum_{\alpha}  
\sigma_{\alpha}+ \frac{1}{2}\sum_{\alpha\beta}p_{\alpha\beta}
\sigma_{\alpha} \sigma_{\beta} $
where $\sigma_\alpha=s_\alpha^{(1)}+\cdots+s_\alpha^{(l)}$
denote $nl$ replicated 
spins, and tr$_s^{(l)}$ denotes the sum over $s_\alpha^{(j)}=\pm 1$.

This expression is quite rich, and embodies the effect of clustering. 
Indeed, by expanding the logarithm in powers of $q_{\alpha\neq\beta}$ 
one observes an effective coupling 
$\hat J_{{\rm eff}}(k)=\hat J(k)/(1-c\beta\hat J(k)q_d)$,
due to the presence of the diagonal elements 
$q_{\alpha\alpha}\equiv q_d(c,T)<1$. 
If $\hat J$ is peaked at some $k$, $\hat J_{{\rm eff}}(k)$ will be 
peaked much stronger, thus  exhibiting clustering effects.
When $\hat J(k)=J_0$ for $k_0<k<k_1$, while vanishing elsewhere,
one considers the long range, oscillating interaction 
$J_{\rm eff}(r)\sim (k_0\cos{k_0r}-k_1\cos{k_1r})/r^2$ at large $r$.
In the scaling limit $k_1-k_0\sim c\to 0$, mean field becomes exact.
Eq. (\ref{freeenergy}) then has as limit the Hopfield model and the
Sherrington-Kirkpatrick-model.~\cite{ThMNsite}

>From  eq. (\ref{freeenergy}) a Ginzburg-Landau  free energy
can be derived. 
Omitting the paramagnetic background, eliminating 
the $q$'s and fluctuations of $p_d\equiv p_{\alpha\alpha}$, 
and denoting $p_{\alpha\beta}$ again by $q_{\alpha\beta}$, 
we end up with 
\begin{eqnarray}
\beta F_n&=&-\frac{h^2}{2}\sum_{\alpha\beta}q_{\alpha\beta}
 -\frac{\tau}{2} \sum_{\alpha} (q^2)_{\alpha\alpha}-\frac{w}{6}
\sum_{\alpha}(q^3)_{\alpha\alpha} \nonumber\\
&-&\frac{y_1}{8}\sum_{\alpha\beta}q^4_{\alpha\beta} 
-\frac{y_2}{8}\sum_{\alpha\beta\gamma}q^2_{\alpha\beta}q^2_{\alpha\gamma}
  -\frac{y_3}{8}\sum_{\alpha}(q^4)_{\alpha\alpha}
\label{GLfe} 
\end{eqnarray}
where now $q_{\alpha\alpha}=0$ and 
$h^2=\beta^2 H^2\mu_2$.
The prefactor of the quadratic term, 
$\tau=(\mu_{22}-T^2/cJ_2)/2$,
vanishes at the 
spin glass 
temperature
 $T_g(c)\equiv \sqrt{cJ_2\mu_{22}}$. Furthermore,
$w=\mu_{222}+T^3 J_3/(c J_2^3)$, $y_1=3\mu_{2222}/2+\mu_{44}/6-\mu_{422}$,
$y_3=\mu_{2222}+T^4(J_2J_4-2J_3^2)/(cJ_2^5)$, and a similar 
expression for $y_2$.
We introduced the moments of the effective coupling
$J_l=\int\frac{{\rm d}^3k}{(2\pi)^3}
[\hat{J}_{\rm eff}( k)]^l$ and the spin-moments 
\begin{equation}
\mu_{k_1 \ldots k_j}\equiv\sum_{l=1}^\infty\gamma_l 
\frac{m_{k_1}^{(l)}}{m_0^{(l)}}\cdots \frac{m_{k_j}^{(l)}}{m_0^{(l)}}
\end{equation}
where $m_k^{(l)}={\rm tr}_\sigma 
 \sigma^k\exp(p_d\sigma^2/2)$ with $\sigma=s^{(1)}+\cdots+s^{(l)}$.

The paramagnetic
behavior is coded in the parameters $p_d$ and $q_d$, that satisfy the 
coupled mean field equations $p_d=\beta J_1$ and $q_d=\mu_2$.
All information on clustering
is contained in $\tau$, $w$, and the $y$'s, so in 
the $\mu$'s and the $J_l$.  
In the limit $c\to 0$ the $\mu$'s go to unity and for $T\sim \sqrt{c}$
the $J_3$ and $J_4$ terms vanish, so that one recovers
the Ginzburg-Landau free energy of the SK-model. 
The important factors then are $w=1$, $y_1=2/3$, while the 
values of $y_2$ $(=-2)$ and $y_3$ $(=1)$ are irrelevant.
When following the transition line $T=T_g(c)$ in the $c-T$ 
phase diagram as function of $c$, it is seen that the higher
$\mu$'s are rapidly oscillating functions. 
For instance, if $J_3/J_2^{3/2}\approx$$J_4/J_2^2\approx 0$, 
then $y_1$ changes sign 
at $c=2.7\%$ and at $c=4.3\%$, while $w$ becomes negative at $6.7\%$. 

Based on these observations we are led to assume that the relevant
physics near the phase transition(s) is still contained in the GL 
free energy (\ref{GLfe}). However, 
there is no reason to assume that $w$ and $y_1$ will always be positive.
(A sign change of $y_1$ occurs also in a Potts glass.~\cite{GKS})
Given the type of the lattice and the values of the spin-spin couplings, the
$c-T$ phase diagram may exhibit a limited number of
special points ($c_\ast,T_\ast$)
where either $w$ or $y_1$ vanishes, and new phase boundaries originate.

When the $q_{\alpha\beta}$ are expressed in the Parisi order parameter
function $q(x)$, one obtains the following free energy:
\begin{eqnarray}\label{F cont with y2,y3}
&\beta& F = \int^1_0 {\rm d}x\left\{\frac{h^2}{2}q(x)+\frac{\tau}{2}q^2(x)-
\frac{w}{3}q(x)T(x)\right.\\
&+&\left.\frac{y_1}{8}q^4(x)-\frac{y_2+y_3}{8}q^2(x)\int^1_0{\rm d}y
q^2(y) +\frac{y_3}{2}T^2(x)\right\}\nonumber
\end{eqnarray}      
with $T(x)=xq^2(x)/2+q(x)\int^1_x{\rm d}yq(y)+
\int_0^x{\rm d}yq^2(y)/2$.

We first investigate the region where $w$ goes through zero
($-1\ll w\ll 1$) while $y_1>0$ is fixed. In figure 1 we depict 
a fictitious phase diagram with such a situation.
On the side where $w>0$ one has the well known spin glass solution of the
Parisi type, as depicted in figure 2a.
The interesting domain is $w<0$ and $\tau\sim w^2$, since $y_2$ and $y_3$ 
become relevant.
In order to find an acceptable solution we assume that $y_3<-y_1$ so 
that $\alpha\equiv \sqrt{-y_1/y_3}<1$. At $h=0$
the spin glass order parameter function
\begin{equation}
 q(x)=\frac{w\sqrt{y_1+y_3x_1^2}}{3(y_1+y_3x_1)}\,\,
\frac{x}{\sqrt{y_1+y_3x^2}}
\end{equation}
has plateau value $q_1=q(x_1)$, determined by 
\begin{equation}\label{d1/dx1 x=x1}
\tau=wq_1-\frac{3}{2}(y_1+y_3)q_1^2+\frac{y_2}{2}(1-x_1)q_1^2
+\frac{y_2}{2}I_2
\end{equation}
where $I_2=\int_0^{x_1}dy q^2(y)$. 
The solution is physically acceptable as soon as $y_2$ exceeds a certain 
bound and exists for parameters such that $x_1$ ranges 
from $x_1=0$ up to $x_1=\alpha$. For $(c,T)$ such that
$x_1\to \alpha$ the solution squeezes and becomes a 1RSB solution
with the lower plateau at $q_0=0$ (in zero field), see figure 2b.

For $w<0$ 1RSB solutions are present in a whole domain.
In general, a 1RSB occurs in two shapes, static and dynamic. The static case 
 describes  physics on exponentially large time scales where the
system can overcome the free energy barriers between pure states. 
Here one maximizes the free energy  wrt $x_1$, which yields
 the plateau value reads 
\begin{eqnarray}
q_1^{\rm g} & = & 
\frac{wx_1}{\frac{3}{2}y_1+3y_3x_1\left(1-\frac{1}{2}x_1\right)}
\label{q1 1RSB stat} 
\end{eqnarray}
It sets in from $x_1=1$ as a first order phase 
transition without latent heat at temperature
\begin{equation}
T_g^{\rm 1RSB}=T_g(c)-\tau_g \equiv 
T_g(c)+\frac{w^2}{9|y_1+y_3|}
\end{equation}
Whereas the transition from paramagnet 
to spin glass has a continuous specific heat, the analogy 
to real glasses makes us expect that (also beyond mean field) 
the specific heat jumps downwards at the transition PM$\to$1RSB. 
Both the SG and 1RSB phases will exhibit a difference
between field cooled and zero field  cooled susceptibilities.

In mean field the metastable states have infinite lifetime. 
Therefore the dynamical 1RSB equations
lead to a sharp phase transition at temperature $T_c>T_g$.
The  thermodynamics of this dynamical transition is uncommon.~\cite{maxmin}
The entropy of the frozen state is much below the paramagnetic one.
A crucial role is played by 
the complexity (configurational entropy), which is extensive.
This scenario explains thermodynamically why the dynamical glass
transition takes place: the system just goes to the available state 
with lowest free energy.~\cite{complexity} 
Beyond mean field the dynamical aspects are
reflected in the dependence on the cooling rate.

For a dynamical 1RSB-phase the $q_1$-plateau is marginally stable and
equal to 
\begin{equation}
q_1^c  =  \frac{wx_1}{2y_1+y_3x_1(3-x_1)}
\label{q1 1RSB marg}
\end{equation}
This dynamical solution sets in at a larger temperature
\begin{equation} 
T_c^{{\rm 1RSB}}=T_g(c)-\tau_c\equiv 
T_g(c)+\frac{w^2}{8|y_1+y_3|}
\end{equation}
Both the static and dynamical  solutions exist down to
\begin{equation}
T_{sg}(w) =  T_g(c)-\tau_{sg} \equiv 
T_g(c)-
\frac{w^2}{6y_3}\left(1+\frac{y_2}{3y_3(1-\alpha)}\right)
\label{sigma_c}
\end{equation}  
This is exactly the line where, coming from positive $w$,
the SG solution gets squeezed into a 1RSB solution. 
The full phase diagram  is depicted in figure 3. 

Next we consider the situation where $y_1$ goes through zero,
while $w>0$ is fixed. (In case $w<0$ the system will already 
have undergone a non-perturbative first order transition at
some negative $\tau$.)
One now expects a transition from a spin glass phase phase ($y_1>0$)
to a replica symmetric or Edwards-Anderson (EA) phase ($y_1<0$).
In the EA phase there is no difference between 
field cooled and zero field  cooled susceptibility.

As it was the case for Parisi's solution of the SK-model, the values 
of $y_2$ and $y_3$ are now irrelevant.
However,  
higher order replica symmetry breaking terms will become relevant.
All fifth order terms have been considered for the above model.
The most dangerous one is
$-(y_5/8)\sum_{\alpha\beta}q^3_{\alpha\beta}(q^2)_{\alpha\beta}$
with $y_5=6\mu_{22222}-4\mu_{4222}+\frac{2}{3}\mu_{442}$.
(For SK: $y_5=8/3$).  
We can absorb this term in our previous free energy
 using the saddle point equation 
$(q^2)_{\alpha\beta}\approx-2\tau q_{\alpha\beta}/w$,
which amounts to replacing $y_1$ by $\tilde y_1=y_1-2\tau y_5/w$. 
The most dangerous sixth order term is 
$-( y_6/6)\sum_{\alpha\beta}q^6_{\alpha\beta}$, where
$ y_6=\frac{15}{4}\mu_{222222}-\frac{15}{4}\mu_{42222}
+\frac{15}{16}\mu_{4422}+\frac{1}{4}\mu_{6222}
-\frac{1}{8}\mu_{642}+\frac{1}{240}\mu_{66}$
($ y_6=16/15$ for SK).

\begin{figure}
\centering
\begin{picture}(0,0)%
\includegraphics{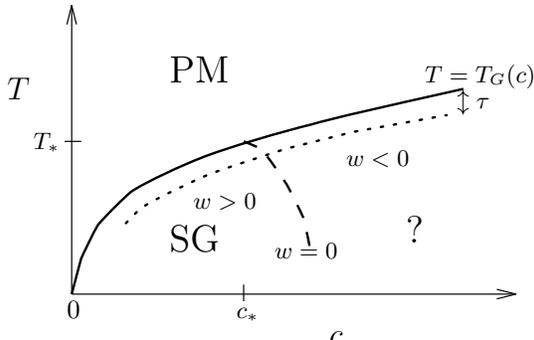}%
\end{picture}%
\setlength{\unitlength}{0.012500in}
\begingroup\makeatletter\ifx\SetFigFont\undefined
\def\x#1#2#3#4#5#6#7\relax{\def\x{#1#2#3#4#5#6}}%
\expandafter\x\fmtname xxxxxx\relax \def\y{splain}%
\ifx\x\y   
\gdef\SetFigFont#1#2#3{%
  \ifnum #1<17\tiny\else \ifnum #1<20\small\else
  \ifnum #1<24\normalsize\else \ifnum #1<29\large\else
  \ifnum #1<34\Large\else \ifnum #1<41\LARGE\else
     \huge\fi\fi\fi\fi\fi\fi
  \csname #3\endcsname}%
\else
\gdef\SetFigFont#1#2#3{\begingroup
  \count@#1\relax \ifnum 25<\count@\count@25\fi
  \def\x{\endgroup\@setsize\SetFigFont{#2pt}}%
  \expandafter\x
    \csname \romannumeral\the\count@ pt\expandafter\endcsname
    \csname @\romannumeral\the\count@ pt\endcsname
  \csname #3\endcsname}%
\fi
\fi\endgroup
\begin{picture}(214,142)(40,638)
\put( 40,741){\makebox(0,0)[lb]{\smash{\SetFigFont{12}{14.4}{rm}$T$}}}
\put(225,737){\makebox(0,0)[lb]{\smash{\SetFigFont{9}{10.8}{rm}
$\updownarrow \tau$}}}
\put(215,748){\makebox(0,0)[lb]{\smash{\SetFigFont{9}{10.8}{rm}$T=T_G(c)$}}}
\put(118,695){\makebox(0,0)[lb]{\smash{\SetFigFont{9}{10.8}{rm}$w>0$}}}
\put( 51,720){\makebox(0,0)[lb]{\smash{\SetFigFont{9}{10.8}{rm}$T_{\ast}$}}}
\put(182,713){\makebox(0,0)[lb]{\smash{\SetFigFont{9}{10.8}{rm}$w<0$}}}
\put(207,681){\makebox(0,0)[lb]{\smash{\SetFigFont{14}{16.8}{rm}?}}}
\put( 65,649){\makebox(0,0)[lb]{\smash{\SetFigFont{10}{12.0}{rm}0}}}
\put(108,677){\makebox(0,0)[lb]{\smash{\SetFigFont{14}{16.8}{rm}SG}}}
\put(108,748){\makebox(0,0)[lb]{\smash{\SetFigFont{14}{16.8}{rm}PM}}}
\put(175,638){\makebox(0,0)[lb]{\smash{\SetFigFont{12}{14.4}{rm}$c$}}}
\put(152,674){\makebox(0,0)[lb]{\smash{\SetFigFont{9}{10.8}{rm}$w=0$}}}
\put(136,649){\makebox(0,0)[lb]{\smash{\SetFigFont{9}{10.8}{rm}$c_{\ast}$}}}
\end{picture}
\caption{ $c-T$ phase diagram for a fictitious system with 
a line $w(c,T)=0$. PM=paramagnet; SG=spin glass.}
\label{fig1}
\end{figure}

The interesting region is where the $q_{\alpha\beta}^4$ term
is of same order of magnitude as the $q_{\alpha\beta}^6$ term.
This occurs when $y_4\equiv\tilde y_1w^2/2\tau^2$ is of order unity.
At fixed small positive $\tau$ we now follow the system by changing $y_4$.
We thus vary $c$ and $T$ over a line at fixed 
distance $\tau$ to the critical line. 
This is indicated by
the dotted line in 
figure~\ref{fig1}, where $w$ should now read $y_1$.
For $y_4\gg 1$ we will have
a standard SG, while for $y_4\ll -1$ there 
is the EA phase.

When $ y_6>0$ is fixed, we find that in between the SG phase 
and the EA phase there is a SG phase with $q_0>0$, although there
is no external field. Coming from the SG-phase, $q_0$ starts to
become non-zero at $y_4=0^{-}$. For $y_4\to -2 y_6$
replica symmetry is restored since
$q_0$ approaches $q_1$.
The $y_1-\tau$ phase diagram for the case $ y_6>0$ is shown 
in figure~\ref{fig4}. As it is the case with the AT-line in a field,
the transition EA $\to$ SG $(q_0\neq 0)$ may very well 
be smeared beyond mean field. 

When $y_6<0$ we find a new, discontinuous order parameter function, 
that we call SG ${ III}$:
$q(x)=q_c(x)$ for $x\le x_1$, while $q(x)=q_1>q_c(x_1)$ for $x>x_1$,
see figure 2c. 
As for static 1RSB solutions, the plateau has stable fluctuations.
Coming from the EA-phase, SG $III$ sets in with $x_1=0$, 
leading to irreversibility.
With respect to the EA-phase, the SG ${III}$ phase
has a smaller replica free energy with 
a discontinuous slope. There occurs a static first order transition
without latent heat but with a discontinuity
in the specific heat, as usual for glasses.

At $y_4=10|y_6|$ the discontinuity of $q(x)$ disappears and the standard
SG solution takes over, see figure ~\ref{fig5}. 

There are also other solutions with free energy between the 
ones of the EA and the SG $III$ states.
At $y_4=|y_6|$ a 1RSB solution with marginal 
lower plateau occurs, as in a Potts glass.~\cite{maxmin}
Now the breakpoint sets in from $x_1=0$.
This 1RSB solution becomes unstable at $y_4=3|y_6|$, where
 the $q_0$ plateau is lifted and a  foot grows near $x=0$. We call this 
the SG $IV$ solution, see figure 2d.
Like the SG $III$ it exists up to $y_4=10|y_6|$,
 where the SG $IV$ discontinuity
disappears and it matches the standard SG solution (see Fig. 2a).
In analogy with the marginal 1RSB solution,
 we anticipate that this 1RSB-SG ${IV}$ traject is the one that occurs 
in dynamics.

Also in the standard region where $w$ and $y_1$ are still positive 
some clustering effects occur. Consider
the slope of the field-cooled susceptibility 
$\chi_{FC}=\beta(1-\int_0^1dx q(x))$. 
At $T_g^{-}$ one has $d\chi_{FC}/dT=-T_g^{-2}+(wT_g)^{-1}d\tau/dT$.
In mean field models with $\infty$RSB $\chi_{FC}$ is usually
constant below $T_g$, so these two terms cancel.  There
does not seem to be a general reason for this. Experimentally,
the values in the SG-phase are usually lower than at $T_g$.
However, in the mechanically milled amorphous Co$_2$Ge spin glass 
of Zhou and Bakker, that has about 67$\%$ of magnetic atoms,
one expects large clustering effects. Indeed,
$\chi_{FC}$ is monotonically decreasing with $T$.~\cite{ZB}
Both these phenomena can be explained by our formula.

So far our results concern mainly mean field. 
Whether or not fluctuations change them qualitatively is unknown.

In conclusion, we have proposed a Ginzburg-Landau free energy for 
site-disordered spin glasses. It is motivated that the prefactors
of the cubic and quartic terms can have zeroes. From these points
new transition lines originate. We find spin glass phases of the
Parisi type ($\infty$RSB), with 1RSB, without RSB (EA-phases), and
of  new types, the SG ${\em III}$ and SG $IV$ phases. For the latter
phases the dynamics will be of new nature. 

\acknowledgments
The authors thank J.A. Mydosh, G. Parisi, and D. Lancaster
 for discussion and J.A. M. also for a critical reading of the manuscript.
Th.M. N.'s research was made possible by the Royal Netherlands Academy 
of Arts and Sciences (KNAW).

\begin{figure}
\centering
\begin{picture}(0,0)%
\includegraphics{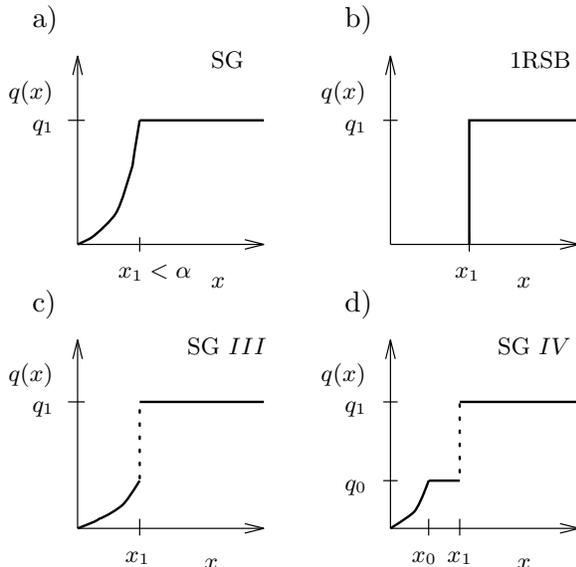}%
\end{picture}%
\setlength{\unitlength}{0.012500in}%
\begingroup\makeatletter\ifx\SetFigFont\undefined
\def\x#1#2#3#4#5#6#7\relax{\def\x{#1#2#3#4#5#6}}%
\expandafter\x\fmtname xxxxxx\relax \def\y{splain}%
\ifx\x\y   
\gdef\SetFigFont#1#2#3{%
  \ifnum #1<17\tiny\else \ifnum #1<20\small\else
  \ifnum #1<24\normalsize\else \ifnum #1<29\large\else
  \ifnum #1<34\Large\else \ifnum #1<41\LARGE\else
     \huge\fi\fi\fi\fi\fi\fi
  \csname #3\endcsname}%
\else
\gdef\SetFigFont#1#2#3{\begingroup
  \count@#1\relax \ifnum 25<\count@\count@25\fi
  \def\x{\endgroup\@setsize\SetFigFont{#2pt}}%
  \expandafter\x
    \csname \romannumeral\the\count@ pt\expandafter\endcsname
    \csname @\romannumeral\the\count@ pt\endcsname
  \csname #3\endcsname}%
\fi
\fi\endgroup
\begin{picture}(240,239)(40,560)
\put( 125,770){\makebox(0,0)[lb]{\smash{\SetFigFont{9}{10.8}{rm}SG}}}
\put( 250,770){\makebox(0,0)[lb]{\smash{\SetFigFont{9}{10.8}{rm}1RSB}}}
\put( 115,650){\makebox(0,0)[lb]{\smash{\SetFigFont{9}{10.8}
{rm}SG$\,\,III$}}}
\put( 245,650){\makebox(0,0)[lb]{\smash{\SetFigFont{9}{10.8}
{rm}SG$\,\,IV$}}}
\put( 40,757){\makebox(0,0)[lb]{\smash{\SetFigFont{9}{10.8}{rm}$q(x)$}}}
\put( 86,682){\makebox(0,0)[lb]{\smash{\SetFigFont{9}{10.8}{rm}$x_1<\alpha$}}}
\put(125,678){\makebox(0,0)[lb]{\smash{\SetFigFont{9}{10.8}{rm}$x$}}}
\put( 50,744){\makebox(0,0)[lb]{\smash{\SetFigFont{9}{10.8}{rm}$q_1$}}}
\put( 50,787){\makebox(0,0)[lb]{\smash{\SetFigFont{11}{13.2}{rm}a)}}}
\put( 40,639){\makebox(0,0)[lb]{\smash{\SetFigFont{9}{10.8}{rm}$q(x)$}}}
\put(122,560){\makebox(0,0)[lb]{\smash{\SetFigFont{9}{10.8}{rm}$x$}}}
\put( 89,563){\makebox(0,0)[lb]{\smash{\SetFigFont{9}{10.8}{rm}$x_1$}}}
\put( 50,626){\makebox(0,0)[lb]{\smash{\SetFigFont{9}{10.8}{rm}$q_1$}}}
\put( 50,668){\makebox(0,0)[lb]{\smash{\SetFigFont{11}{13.2}{rm}c)}}}
\put(181,668){\makebox(0,0)[lb]{\smash{\SetFigFont{11}{13.2}{rm}d)}}}
\put(171,757){\makebox(0,0)[lb]{\smash{\SetFigFont{9}{10.8}{rm}$q(x)$}}}
\put(253,678){\makebox(0,0)[lb]{\smash{\SetFigFont{9}{10.8}{rm}$x$}}}
\put(227,682){\makebox(0,0)[lb]{\smash{\SetFigFont{9}{10.8}{rm}$x_1$}}}
\put(181,744){\makebox(0,0)[lb]{\smash{\SetFigFont{9}{10.8}{rm}$q_1$}}}
\put(181,787){\makebox(0,0)[lb]{\smash{\SetFigFont{11}{13.2}{rm}b)}}}
\put(171,639){\makebox(0,0)[lb]{\smash{\SetFigFont{9}{10.8}{rm}$q(x)$}}}
\put(253,560){\makebox(0,0)[lb]{\smash{\SetFigFont{9}{10.8}{rm}$x$}}}
\put(181,626){\makebox(0,0)[lb]{\smash{\SetFigFont{9}{10.8}{rm}$q_1$}}}
\put(224,563){\makebox(0,0)[lb]{\smash{\SetFigFont{9}{10.8}{rm}$x_1$}}}
\put(209,563){\makebox(0,0)[lb]{\smash{\SetFigFont{9}{10.8}{rm}$x_0$}}}
\put(181,593){\makebox(0,0)[lb]{\smash{\SetFigFont{9}{10.8}{rm}$q_0$}}}
\end{picture}
\caption{Shapes of the spin glass order parameter function.
a) standard form for infinite order replica symmetry breaking;
b) one step replica symmetry breaking solution. 
c) the discontinuous SG {\it III} function ;
d) the SG {\it IV} function. }
\label{fig2}
\end{figure}

\begin{figure}
\centering
\begin{picture}(0,0)%
\includegraphics{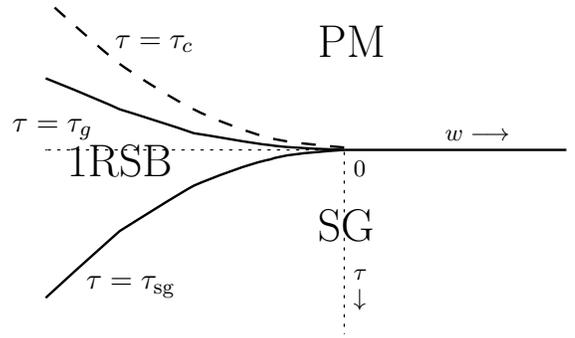}%
\end{picture}%
\setlength{\unitlength}{0.012500in}
\begingroup\makeatletter\ifx\SetFigFont\undefined
\def\x#1#2#3#4#5#6#7\relax{\def\x{#1#2#3#4#5#6}}%
\expandafter\x\fmtname xxxxxx\relax \def\y{splain}%
\ifx\x\y   
\gdef\SetFigFont#1#2#3{%
  \ifnum #1<17\tiny\else \ifnum #1<20\small\else
  \ifnum #1<24\normalsize\else \ifnum #1<29\large\else
  \ifnum #1<34\Large\else \ifnum #1<41\LARGE\else
     \huge\fi\fi\fi\fi\fi\fi
  \csname #3\endcsname}%
\else
\gdef\SetFigFont#1#2#3{\begingroup
  \count@#1\relax \ifnum 25<\count@\count@25\fi
  \def\x{\endgroup\@setsize\SetFigFont{#2pt}}%
  \expandafter\x
    \csname \romannumeral\the\count@ pt\expandafter\endcsname
    \csname @\romannumeral\the\count@ pt\endcsname
  \csname #3\endcsname}%
\fi
\fi\endgroup
\begin{picture}(237,140)(40,637)
\put(187,649){\makebox(0,0)[lb]{\smash{\SetFigFont{9}{10.8}{rm}$\downarrow$}}}
\put(187,660){\makebox(0,0)[lb]{\smash{\SetFigFont{9}{10.8}{rm}$\tau$}}}
\put(187,703){\makebox(0,0)[lb]{\smash{\SetFigFont{9}{10.8}{rm}0}}}
\put( 83,757){\makebox(0,0)[lb]{\smash{\SetFigFont{11}{13.2}{rm}
$\tau=\tau_c$}}}
\put(172,753){\makebox(0,0)[lb]{\smash{\SetFigFont{17}{20.4}{rm}PM}}}
\put(222,718){\makebox(0,0)[lb]{\smash{\SetFigFont{9}{10.8}{rm}
$w\longrightarrow $}}}
\put(172,676){\makebox(0,0)[lb]{\smash{\SetFigFont{17}{20.4}{rm}SG}}}
\put( 71,656){\makebox(0,0)[lb]{\smash{\SetFigFont{11}{13.2}{rm}
$\tau=\tau_{\rm sg}$}}}
\put( 67,703){\makebox(0,0)[lb]{\smash{\SetFigFont{17}{20.4}{rm}1RSB}}}
\put( 40,722){\makebox(0,0)[lb]{\smash{\SetFigFont{11}{13.2}{rm}
$\tau=\tau_g$}}}
\end{picture}
\vspace{.5cm}
\caption{ $\tau-w$ phase diagram for a system with 
$y_1>0$, $y_3<-y_1$ and $y_2$ sufficiently positive; with $w$ increasing
from right to left
 it may appear in Fig. 1 around the point ($c_\ast$, $T_\ast$).
The full (dashed) lines are static  (dynamical) transition lines.}
\label{fig3}  
\end{figure}

\begin{figure}
\centering
\begin{picture}(0,0)%
\includegraphics{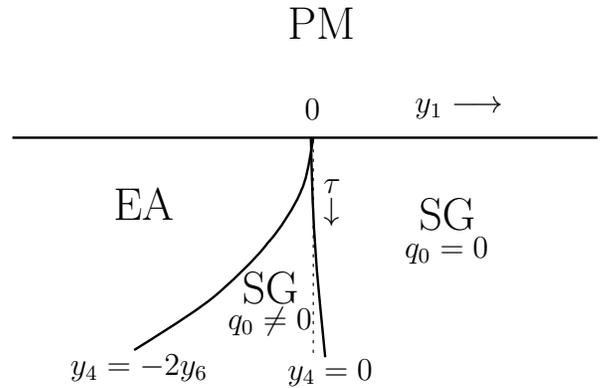}%
\end{picture}%
\setlength{\unitlength}{0.012500in}%
\begingroup\makeatletter\ifx\SetFigFont\undefined
\def\x#1#2#3#4#5#6#7\relax{\def\x{#1#2#3#4#5#6}}%
\expandafter\x\fmtname xxxxxx\relax \def\y{splain}%
\ifx\x\y   
}\gdef\SetFigFont#1#2#3{%
  \ifnum #1<17\tiny\else \ifnum #1<20\small\else
  \ifnum #1<24\normalsize\else \ifnum #1<29\large\else
  \ifnum #1<34\Large\else \ifnum #1<41\LARGE\else
     \huge\fi\fi\fi\fi\fi\fi
  \csname #3\endcsname}%
\else
\gdef\SetFigFont#1#2#3{\begingroup
  \count@#1\relax \ifnum 25<\count@\count@25\fi
  \def\x{\endgroup\@setsize\SetFigFont{#2pt}}%
  \expandafter\x
    \csname \romannumeral\the\count@ pt\expandafter\endcsname
    \csname @\romannumeral\the\count@ pt\endcsname
  \csname #3\endcsname}%
\fi
\fi\endgroup
\begin{picture}(245,164)(44,616)
\put(208,669){\makebox(0,0)[lb]{\smash{\SetFigFont{12}{14.4}{rm}$q_0=0$}}}
\put(140,650){\makebox(0,0)[lb]{\smash{\SetFigFont{17}{20.4}{rm}SG}}}
\put( 86,685){\makebox(0,0)[lb]{\smash{\SetFigFont{17}{20.4}{rm}EA}}}
\put(159,761){\makebox(0,0)[lb]{\smash{\SetFigFont{17}{20.4}{rm}PM}}}
\put(208,730){\makebox(0,0)[lb]{\smash{\SetFigFont{12}{14.4}{rm}
$y_1\longrightarrow$}}}
\put(134,639){\makebox(0,0)[lb]{\smash{\SetFigFont{12}{14.4}{rm}$q_0\neq 0$}}}
\put(166,727){\makebox(0,0)[lb]{\smash{\SetFigFont{12}{14.4}{rm}0}}}
\put( 68,620){\makebox(0,0)[lb]{\smash{\SetFigFont{12}{14.4}{rm}$y_4=-2y_6$}}}
\put(214,681){\makebox(0,0)[lb]{\smash{\SetFigFont{17}{20.4}{rm}SG}}}
\put(174,685){\makebox(0,0)[lb]{\smash{\SetFigFont{12}{14.4}{rm}$\downarrow$}}}
\put(174,696){\makebox(0,0)[lb]{\smash{\SetFigFont{12}{14.4}{rm}$\tau$}}}
\put(159,618){\makebox(0,0)[lb]{\smash{\SetFigFont{12}{14.4}{rm}$y_4=0$}}}
\end{picture}
\vspace{.5cm}
\caption{$y_1-\tau$ phase diagram for $w>0$, $ y_6>0$. 
The 
function $q(x)$ in the SG phase 
is drawn in figure 2.a for the case $q_0=0$.
In the EA-phase $q(x)$ is constant (no RSB).}
\label{fig4}
\vspace{.5cm}
\end{figure}

\begin{figure}
\centering
\begin{picture}(0,0)%
\includegraphics{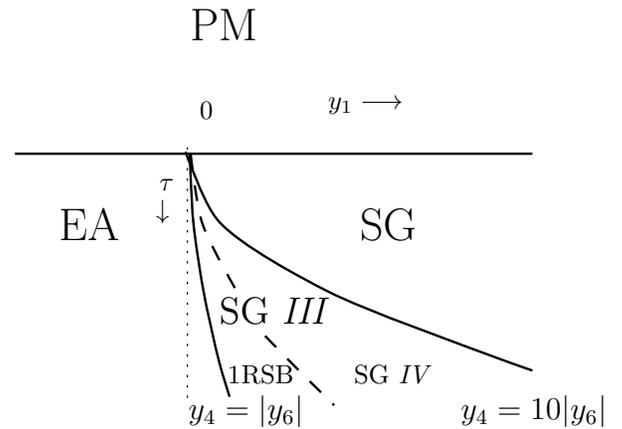}%
\end{picture}%
\setlength{\unitlength}{0.012500in}%
\begingroup\makeatletter\ifx\SetFigFont\undefined
\def\x#1#2#3#4#5#6#7\relax{\def\x{#1#2#3#4#5#6}}%
\expandafter\x\fmtname xxxxxx\relax \def\y{splain}%
\ifx\x\y   
\gdef\SetFigFont#1#2#3{%
  \ifnum #1<17\tiny\else \ifnum #1<20\small\else
  \ifnum #1<24\normalsize\else \ifnum #1<29\large\else
  \ifnum #1<34\Large\else \ifnum #1<41\LARGE\else
     \huge\fi\fi\fi\fi\fi\fi
  \csname #3\endcsname}%
\else
\gdef\SetFigFont#1#2#3{\begingroup
  \count@#1\relax \ifnum 25<\count@\count@25\fi
  \def\x{\endgroup\@setsize\SetFigFont{#2pt}}%
  \expandafter\x
    \csname \romannumeral\the\count@ pt\expandafter\endcsname
    \csname @\romannumeral\the\count@ pt\endcsname
  \csname #3\endcsname}%
\fi
\fi\endgroup
\begin{picture}(252,184)(44,636)
\put(102,724){\makebox(0,0)[lb]{\smash{\SetFigFont{10}{12.0}{rm}
$\downarrow$}}}
\put( 65,715){\makebox(0,0)[lb]{\smash{\SetFigFont{17}{20.4}{rm}EA}}}
\put(107,736){\makebox(0,0)[lb]{\smash{\SetFigFont{10}{12.0}{rm}$\tau$}}}
\put(174,770){\makebox(0,0)[lb]{\smash{\SetFigFont{10}{12.0}{rm}
$y_1\longrightarrow$}}}
\put(127,680)  
{\makebox(0,0)[lb]{\smash{\SetFigFont{14}{16.8}{rm}
SG {\em III}}}}
\put(132,655)  
{\makebox(0,0)[lb]{\smash{\SetFigFont{10}{16.8}{rm}
1RSB}}}
\put(185,655)  
{\makebox(0,0)[lb]{\smash{\SetFigFont{10}{16.8}{rm}
SG {\em IV}}}}
\put(120,799){\makebox(0,0)[lb]{\smash{\SetFigFont{17}{20.4}{rm}PM}}}
\put(124,766){\makebox(0,0)[lb]{\smash{\SetFigFont{10}{12.0}{rm}0}}}
\put(229,640){\makebox(0,0)[lb]{\smash{\SetFigFont{12}{14.4}{rm}
$y_4=10 |y_6|$}}}
\put(191,715){\makebox(0,0)[lb]{\smash{\SetFigFont{17}{20.4}{rm}SG}}}
\put(115,640)
{\makebox(0,0)[lb]{\smash{\SetFigFont{12}{14.4}{rm}
$y_4=|y_6|$}}}
\end{picture}
\vspace{.3cm}
\caption{$y_1-\tau$ phase diagram for $w>0$, $ y_6<0$.
In the SG{\em III} phase $q(x)$ is as in figure 2c.
Dynamically this phase splits up in a 1RSB phase and a SG {\rm IV} phase
, see fig. 2b,d. }
\label{fig5}
\end{figure}


\references
\bibitem{Mydosh} 
J.A. Mydosh, {\em Spin Glasses, an experimental introduction} 
(Taylor and Francis, London, 1993)
\bibitem{ThMNsite} Th.M. Nieuwenhuizen, Europhys.Lett. {\bf 24} 
(1993) 797
\bibitem{IMISI} M. Iguchi, F. Matsubara, T. Iyota, 
T. Shirakura, and S. Inawashiro, Phys. Rev. B {\bf 47} (1993) 2648
\bibitem{DHSS} 
In the approach of ref. ~\cite{ThMNsite} 
this follows immediately from the onset of SG ($T_g\sim \sqrt{c}$)
and ferromagnetic phases ($T_F\sim c$) at low $c$.
For a  detailed analysis in a related model, 
see D.S. Dean and D. Lancaster, to appear. For RSB in
renormalization group flows, see
V. Dotsenko, A.B. Harris, D. Sherrington and R.B. Stinchcombe,
J. Phys. A {\bf 28} (1995) 3093.
\bibitem{GKS} D.J. Gross, I. Kanter, and H. Sompolinsky, Phys. Rev. Lett.
{\bf 55} (1985) 304
\bibitem{KT} T.R. Kirkpatrick and D. Thirumalai, 
Phys. Rev. Lett. {\bf 58} (1987) 2091
\bibitem{CHS} A. Crisanti, H. Horner, and H.J. Sommers, Z. Phys.
B {\bf 92} (1993) 257
\bibitem{CK} L. F. Cugliandolo and J. Kurchan, Phys. Rev. Lett.
{\bf 71} (1993) 173
\bibitem{maxmin} Th.M. Nieuwenhuizen, Phys. Rev. Lett. {\bf 74} 
(1995) 3463 
\bibitem{complexity} Th.M. Nieuwenhuizen, 
preprint; cond-mat/9504059
\bibitem{ZB} G.F. Zhou and H. Bakker, 
Phys. Rev. Lett. {\bf 72} (1994) 2290
\end{document}